  \documentclass[12pt]{article}
  \usepackage{times}

  \topmargin - 0.5 cm 
\begin{document}

\baselineskip = 0.80 true cm

\begin{center}
{\large \bf Representation theory and Wigner-Racah algebra}\\ 
% \vspace{0.1cm}
{\large \bf of the SU(2) group in a noncanonical 
basis\footnote{Dedicated to Professor Josef 
Paldus on the occasion of his 70th birthday.}} 
\end{center}

\vspace{0.5cm}

\begin{center}
{\bf M.R.~Kibler}
\end{center}

\begin{center}
{Institut de Physique Nucl\'eaire de Lyon}\\
{IN2P3-CNRS et Universit\'e Claude Bernard Lyon 1}\\
{43 bd du 11 novembre 1918}\\
{69622 Villeurbanne Cedex, France}
\end{center}

 \vspace{2.5cm}

% \noindent {\bf Abstract}
% \noindent {\LARGE \bf Abstract}
% \noindent {\Large \bf Abstract}
  \noindent {\large \bf Abstract}

%  \begin{abstract}

\noindent 
The Lie algebra su(2) of the classical group SU(2) is built from two commuting 
quon algebras for which the deformation parameter is a common root of unity. This 
construction leads to (i) a not very well-known polar decomposition of the 
generators $J_-$ and $J_+$ of the SU(2) group, 
with $J_+ = J_-^{\dagger} = H U_r$ where $H$ is Hermitean and $U_r$ unitary, 
and (ii) an alternative to the $\{ J^2, J_z \}$ quantization scheme, viz., 
the $\{ J^2, U_r \}$ quantization scheme. The representation theory of the 
SU(2) group can be developed in this nonstandard scheme. The key ideas 
for developing the Wigner-Racah 
algebra of the SU(2) group in the $\{ J^2, U_r \}$ scheme are given. In 
particular, some properties of the coupling and recoupling coefficients as well 
as the Wigner-Eckart theorem in the $\{ J^2, U_r \}$ scheme are examined 
in great detail. 

%  \end{abstract}

%  \vfill
%  \thispagestyle{empty}
%  \end{titlepage}

   \newpage

\section{Introduction}
The concepts of symmetry 
(introduced in a group theoretical context in the 1930's), of supersymmetry 
(introduced in a supergroup context in the 1970's) and of deformations 
(introduced in a bi-algebra context in the 1980's) are of paramount importance 
for quantum chemistry and/or quantum physics. These concepts are often used in 
the exploration of dynamical systems as for example the Coulomb system and 
the oscillator system which can be viewed as two paradigms for the study 
of atomic and molecular interactions.$^{1,2}$ In these directions, 
the works$^3$ of Paldus and its 
collaborators on the second quantization method, the 
unitary group approach and its extension by means of Clifford algebras 
proved to be very useful in numerous domains of theoretical 
chemistry.   

In recent years, the use of deformed oscillator algebras proved to be useful
for many applications of quantum mechanics. For instance, one- and two-parameter 
deformations of oscillator algebras and Lie algebras were applied to 
statistical mechanics$^4$ and to molecular and nuclear physics.$^5$ 

It is the purpose of this work to apply deformed 
oscillator algebras or quon algebras to the representation theory 
and the Wigner-Racah algebra of the SU(2) group. The notion of
deformation is very familiar to the theoretician. In this connection, 
quantum mechanics may be considered as a deformation (the deformation 
parameter being the rationalised Planck constant $\hbar$) of 
classical mechanics. In the same vein, relativistic mechanics is, 
to some extent, another deformation 
(with the inverse of the velocity of light $c^{-1}$ as deformation parameter) 
of classical mechanics. The idea of a deformation of an oscillator algebra and 
of a Lie algebra also relies on the introduction of a deformation parameter $q$ 
such that the limiting situation where $q=1$ corresponds to the nondeformed 
algebraic structure. 

The organisation of this paper is as follows. Section 2 is devoted to 
some generalities on the notion of a Wigner-Racah algebra of a finite or 
compact group. In Section 3, we construct the Lie algebra of SU(2) 
from two quon algebras $A_1$ and $A_2$ corresponding to the same deformation 
parameter $q$ taken as a root of unity. Section 4 deals with an alternative 
to the $\{ J^2 , J_z\}$ scheme of SU(2), viz. the $\{ J^2 , U_r\}$ scheme, 
and with the basic elements for the representation theory of SU(2) in this 
scheme. Finally, we develop in Section 5 the Wigner-Racah algebra of SU(2) 
in the $\{ J^2 , U_r\}$ scheme.  

Throughout the present work, we use the notation $[A , B]$ 
for the commutator of $A$ and $B$. As usual, ${z}^*$ 
denotes the complex conjugate of the number $z$ and 
$A^{\dagger}$ stands for the Hermitean conjugate 
of the operator $A$. 

\section{Wigner-Racah algebra of SU(2)}
The mathematical structure of a Wigner-Racah algebra (WRa) 
associated with a group takes its origin in 
the works by Wigner$^6$ on a simply reducible 
group, with emphasis on the ordinary rotation group, 
and by Racah$^7$ on chains of groups of type 
$\mbox{SU} (2 \ell + 1) \supset 
 \mbox{SO} (2 \ell + 1) \supset 
 \mbox{SO} (3)$, mainly with $\ell = 2, 3$.
From a practical point of view, the WRa of a group deals with the
algebraic relations satisfied by its coupling 
and recoupling coefficients. From a more
theoretical point of view, the WRa of a finite or compact 
group can be defined to be the infinite-dimensional 
Lie algebra spanned by the Wigner  unit  operators 
(i.e., the operators whose matrix elements 
are the coupling or Clebsch-Gordan
or Wigner coefficients of the group).$^8$ 

The WRa of the SU(2) group
is well known. It is generally developed in the standard basis
$\left\{ |j  m \rangle : 2j \in {\bf N}, 
           \ m = - j, - j + 1, \cdots, j \right\}$
arising in the simultaneous diagonalization of the Casimir 
operator $J^2$ and of one generator, say $J_z$, of 
SU(2). Besides this basis, there exist 
several other bases. Indeed, any change of basis 
of type
\begin{eqnarray}
| j  \mu \rangle = \sum_{m=-j}^{j} | j  m \rangle \langle jm | j \mu \rangle 
\end{eqnarray}
(where the $(2 j + 1)\times(2 j + 1)$ matrix with elements  
$\langle jm | j \mu \rangle$ is an arbitrary unitary matrix) 
defines another acceptable basis for the WRa of 
SU(2). In this basis, the matrices of the irreducible 
representation classes of SU(2)
take a new form as well as the coupling coefficients 
(and the associated 3-$jm$ symbols).   As a matter of fact, 
the coupling coefficients $(j_1 j_2 m_1 m_2 | j m)$ are simply replaced by 
   \begin{eqnarray} 
(j_1 j_2 \mu_1 \mu_2 | j \mu) = 
\sum_{m_1=-j_1}^{j_1} 
\sum_{m_2=-j_2}^{j_2} 
\sum_{m  =-j  }^{j  }     (j_1 j_2 m_1 m_2 | j m)          \nonumber \\
                   \langle j_1m_1 | j_1 \mu_1 \rangle^* \,
                   \langle j_2m_2 | j_2 \mu_2 \rangle^* \,
                   \langle j  m   | j   \mu   \rangle
   \end{eqnarray}
when passing from the $\left\{ j m   \right\}$ quantization to the 
                      $\left\{ j \mu \right\}$ quantization 
while the recoupling coefficients, 
and the corresponding 3$(n-1)$-$j$ symbols, 
for the coupling of $n$ ($n > 2$) angular momenta 
remain invariant. 

The various bases for  $\mbox{SU}(2)$  may be classified into two types:
group-subgroup type and nongroup-subgroup type. The standard basis corresponds
to a group-subgroup type basis associated with the chain of groups 
 $\mbox{SU}(2) \supset \mbox{U}(1)$. Another 
group-subgroup type basis may be obtained by replacing 
 $\mbox{U}(1)$ by a finite group $G^*$ (generally the double, i.e., 
spinor group, of a point group $G$ of molecular or crystallographic interest). 
Among the  $\mbox{SU}(2) \supset G^*$  bases, we may distinguish: 
the weakly symmetry-adapted bases for which the basis vectors are
eigenvectors of $J^2$ and of the projection operators of $G^*$
(e.g., see Ref.~[9]) and 
the strongly symmetry-adapted bases for which the basis vectors are
eigenvectors of $J^2$ and of an operator defined in the enveloping algebra 
of  SU(2) and invariant under the group $G$ 
(e.g., see Ref.~[10]). We shall 
see that the basis for SU(2) described in the present paper interpolates 
between the group-subgroup type and the nongroup-subgroup type.

\section{A quon realization of the algebra su(2)}
\subsection{Two quon algebras}
The concept of quon takes its origin in the replacement of the 
commutation (sign $-$) and anticommutation (sign $+$) relations
 \begin{eqnarray}
 a_- a_+ \pm a_+ a_- = 1
 \end{eqnarray}
by the relation 
 \begin{eqnarray}
 a_- a_+ - q a_+ a_- = 1
 \end{eqnarray} 
where $q$ is a constant. Following the works in Ref.~[11],
we define two commuting quon algebras $A_i = \{ a_{i-}, a_{i+}, N_i \}$ 
with $i = 1$ and $2$ by
  \begin{equation}
   a_{i-}a_{i+} - qa_{i+}a_{i-} = 1,            \quad  
   \left[ N_i, a_{i\pm} \right] = \pm a_{i\pm}, \quad
   N_i^{\dagger} = N_i
  \end{equation}
  \begin{equation}
  \left( a_{i+} \right)^k = \left( a_{i-} \right)^k = 0  
  \end{equation}
  \begin{equation}
  \forall x_1 \in A_1, \ \forall x_2 \in A_2 : [x_1 , x_2] = 0  
  \end{equation}
where
    \begin{equation}
     q = \exp \left( {2 \pi {\rm i} \over k} \right)  \quad  {\rm with}  \quad  
     k \in {\bf N} \setminus \{ 0,1 \} 
    \end{equation}
Equation (5) corresponds to the {\em \`a la} Arik and 
Coon$^{11}$ relations defining a quon algebra except that, in the present work, 
$q$ is a root of unity instead of being a positive real number. 
The deformation parameter $q$ is the same for each of the algebras $A_1$ 
and $A_2$ so that $A_1$ and $A_2$ can be considered as two copies 
of the same quon algebra. Equation (6) constitutes nilpotency conditions 
which are indeed compatibility relations to account for the fact that 
$q$ is not a positive number (remember $q^k = 1$). Equation (7) reflects 
the commutativity of the algebras $A_1$ and $A_2$. The generators $a_{i \pm}$
and $N_i$ of $A_1$ and $A_2$ are linear operators. As in the classical case 
$q=1$, we say that $a_{i +}$ is a creation operator, $a_{i -}$ an annihilation 
operator and $N_i$ a number operator (with $i = 1 , 2$). However, the operator 
$a_{i +}$ cannot be considered as the adjoint of the operator $a_{i -}$ except 
for $k = 2$ and $k \to \infty$. In contrast, the operator $N_i$ can be taken 
to be a Hermitean operator for any value of $k$ in 
${\bf N} \setminus \{ 0 , 1\}$. It should be observed that $N_i$ is different 
from $a_{i +} a_{i -}$ except for $k = 2$ and $k \to \infty$. Note that the case
$k=2$ ($\Rightarrow q = -1$) corresponds to fermion operators and the case 
$k \to \infty$ ($\Rightarrow q \to 1$) to boson operators. In other words, each 
of the algebras $A_i$ describes fermions for $q=-1$ and bosons for $q=1$ with 
$N_i = a_{i +} a_{i -}$ for fermions and bosons ($i = 1 , 2$).

To close this subsection, let us mention that algebras similar to $A_1$ and $A_2$ with 
$N_1 \equiv N_2$ were introduced by Daoud, Hassouni and Kibler$^{11}$ for defining
$k$-fermions which are, like anyons, objects interpolating between fermions 
(corresponding to $k = 2$) and bosons (corresponding to $k \to \infty$).

\subsection{Representation of the quon algebras}
We can find several Hilbertian representations of the algebras 
$A_1$ and $A_2$. In this work, we take the representation 
of $A_1 \otimes A_2$ defined by the following action 
  \begin{eqnarray}
  a_{1+} |n_1 , n_2) = |n_1 + 1 , n_2),                       \quad  
  a_{1+} |k-1 , n_2) = 0 
  \end{eqnarray}
  \begin{eqnarray}
  a_{1-} |n_1 , n_2) = \left[ n_1   \right]_q |n_1-1 , n_2),  \quad   
  a_{1-} |0   , n_2)   = 0
  \end{eqnarray}
  \begin{eqnarray}
  a_{2+} |n_1 , n_2) = \left[ n_2+1 \right]_q |n_1 , n_2+1),  \quad  
  a_{2+} |n_1 , k-1) = 0
  \end{eqnarray}
  \begin{eqnarray}
  a_{2-} |n_1 , n_2) = |n_1 , n_2 - 1),                       \quad  
  a_{2-} |n_1 , 0) = 0
  \end{eqnarray}
  \begin{eqnarray}
  N_1 |n_1 , n_2) = n_1 |n_1 , n_2),                          \quad  
  N_2 |n_1 , n_2) = n_2 |n_1 , n_2)
  \end{eqnarray}
on a finite (Fock) space 
${\cal F}_k = \{ | n_1 , n_2 ) : n_1 , n_2 = 0, 1, \cdots, k-1 \}$ 
of dimension dim~${\cal F}_k = k^2$. In Eqs.~(10) and (11), we use 
  \begin{eqnarray}
  \left[ x \right]_q = \frac{1-q^x}{1-q}  \quad  {\hbox{for}}  \quad  x \in {\bf R}
  \end{eqnarray}
which yields 
\begin{eqnarray}
[n]_q = 1 + q + \cdots + q^{n-1} \quad {\rm for} \quad n \in N^*
                                 \quad {\rm and} \quad [0]_q = 0 
\end{eqnarray}
as a particular case. We shall also use the $q$-deformed factorial 
defined by 
  \begin{eqnarray}
  \left[ n \right]_q! = 
  \left[ 1 \right]_q 
  \left[ 2 \right]_q \cdots 
  \left[ n \right]_q \quad {\rm for} \quad n \in {\bf N}^*, 
                                     \quad \left[ 0 \right]_q! = 1  
  \end{eqnarray}
so that 
$\left[ n+1 \right]_q! = 
 \left[ n   \right]_q! 
 \left[ n+1 \right]_q$ 
for $n \in {\bf N}$.

The space ${\cal F}_k$ is a unitary space with a scalar product noted 
$( \ | \ )$. The $k^2$ vectors $|n_1 , n_2)$ are taken in a form 
such that    
\begin{eqnarray}
( n_1' , n_2' | n_1 , n_2 ) = \delta (n_1' , n_1 ) \>
                              \delta (n_2' , n_2 )
\end{eqnarray}
(i.e., they constitute an orthonormalized basis of ${\cal F}_k$). The 
space ${\cal F}_k$ turns out to be the direct product 
${\cal F}(1) \otimes {\cal F}(2)$ of two truncated Fock spaces 
${\cal F}(i) = \{ | n_i ) : n_i = 0, 1, \cdots, k-1 \}$ of 
dimension $\dim {\cal F}(i) = k$ ($i=1,2$) corresponding to 
two truncated harmonic oscillators. At this stage, we realize 
why the cases $k=0$ and $k=1$ should be excluded.  
The case $k=1$ would give trivial algebras $A_i$ with 
$a_{i-} = a_{i+} = 0$ ($i=1,2$) and the case $k=0$ 
would lead to a nondefined value of $q$.
 
\subsection{Two important operators}
We now define the two linear operators
  \begin{eqnarray}
  H = {\sqrt {N_1 \left( N_2 + 1 \right) }}
  \end{eqnarray}
and
  \begin{eqnarray}
  U_r  =       \left[ a_{1+} + 
  {\rm e}^{\frac{1}{2} {\rm i} \phi_r}  {1 \over 
  \left[ k-1 \right]_q!} (a_{1-})^{k-1} \right] \left[ a_{2-} + 
  {\rm e}^{\frac{1}{2} {\rm i} \phi_r}  {1 \over 
  \left[ k-1 \right]_q!} (a_{2+})^{k-1} \right]
  \end{eqnarray}
% (formule de U_r)  
where the arbitrary real parameter $\phi_r$ is taken in the form 
  \begin{eqnarray}
  \phi_r = \pi (k-1) r  \quad  {\rm with}  \quad  r \in {\bf R}
  \end{eqnarray}
It is immediate to show that the action of $H$ and $U_r$ on 
${\cal F}_k$ is given by
  \begin{equation}
  H |n_1 , n_2) = {\sqrt{ n_1 (n_2 + 1) } |n_1 , n_2)}
  \quad {\rm for}  \quad n_i = 0, 1, 2, \cdots, k-1 
  \quad {\rm with} \quad i=1,2
  \end{equation}
and
  \begin{equation}
  U_r |n_1 , n_2) = |n_1+1 , n_2-1) 
                  \quad {\rm for} \quad n_1 \not = k-1  
                  \quad {\rm and} \quad n_2 \not = 0
  \end{equation}
  \begin{equation}
  U_r |k-1 , n_2) = {\rm e}^{\frac{1}{2} {\rm i} {\phi}_r} 
  |0 , n_2 - 1)   \quad {\rm for} \quad n_2 \not= 0
  \end{equation}
  \begin{equation}
  U_r |n_1 , 0) = {\rm e}^{\frac{1}{2} {\rm i} {\phi}_r} 
  |n_1 + 1 , k-1) \quad {\rm for} \quad n_1 \not= k-1  
  \end{equation}
  \begin{equation}
  U_r |k-1 , 0) = {\rm e}^{{\rm i} {\phi}_r} 
  |0 , k-1)       \quad {\rm for} \quad n_1 = k-1  
                  \quad {\rm and} \quad n_2 = 0
  \end{equation}

The operators $H$ and $U_r$ satisfy interesting properties. First, it 
is obvious that the operator $H$ is Hermitean. Second, the operator 
$U_r$ is unitary. In addition, the action of $U_r$ on the space ${\cal F}_k$
is cyclic. More precisely, we can check that
\begin{eqnarray}
(U_r)^k = {\rm e}^{{\rm i} {\phi}_r} I
\end{eqnarray}
where $I$ is the identity operator.

From the Schwinger work on angular momentum,$^{12}$ we introduce 
  \begin{eqnarray}
  J = {1 \over 2} \left( n_1+n_2 \right),  \quad  
  M = {1 \over 2} \left( n_1-n_2 \right)  
  \end{eqnarray}
Consequently, we can write 
  \begin{eqnarray} 
  |n_1 , n_2) = |J + M , J-M) 
  \end{eqnarray}  
We shall use the notation 
  \begin{eqnarray} 
  |J M \rangle \equiv |J + M , J-M) 
  \end{eqnarray}
for the vector $|J + M , J-M)$. For a fixed value of $J$,
the label $M$ can take $2J+1$ values $M = -J, -J+1, \cdots, J$.  
Equations (28) and (29) proved to be of central importance 
for the connection between angular momentum and a coupled pair 
of ordinary harmonic oscillators.$^{12}$ We guess here that 
they shall play an important role for connecting the Lie algebra 
of su(2) to a coupled pair of truncated harmonic oscillators. 

For fixed $k$, the maximum value of $J$ is 
\begin{eqnarray}
J = J_{\rm max} = k-1
\end{eqnarray}
and the following value of $J$
\begin{eqnarray}
J = j = \frac{1}{2} (k-1)
\end{eqnarray}
is admissible. For a given value of $k \in {\bf N} \setminus \{0 , 1\}$, 
the $2j+1=k$ vectors $|j m \rangle$ belong to the vector space 
${\cal F}_k$. Let $\varepsilon (j)$ be the subspace of ${\cal F}_k$, 
of dimension $\dim \varepsilon (j) = k$, spanned by the $k$ vectors 
$|j m \rangle$. We can thus associate the space
\begin{eqnarray}
\varepsilon (j) = \{ |j m \rangle : m = -j, -j+1, \cdots, j \}
\end{eqnarray}
for $j = \frac{1}{2}, 1, \frac{3}{2}, \cdots$ to the values 
    $k = 2, 3, 4, \cdots$, respectively. The case $\varepsilon (j=0)$ 
can be seen to correspond to the limiting situation where $k \to \infty$. 
    
    The action of the operators $H$ and $U_r$ on the space $\varepsilon (j)$
can be described by 
  \begin{eqnarray}
  H |j m \rangle = {\sqrt{ (j+m)(j-m+1) }} |j m \rangle
  \end{eqnarray} 
and
  \begin{eqnarray}
  U_r |jm \rangle = \left[ 1 - \delta (m,j) \right] |j m+1 \rangle + \delta(m,j)  
  {\rm e}^{{\rm i} {\phi}_r}
  |j -j \rangle 
  \end{eqnarray}
which are a simple rewriting, in terms of the vectors $|j m \rangle$, 
of Eqs.~(21) and (22)-(25), respectively. The subspace 
$\varepsilon (j)$ of ${\cal F}_k$ is thus stable under $H$ and $U_r$. Furthermore,
the action of the adjoint $U_r^{\dagger}$ of $U_r$ on the space $\varepsilon (j)$ 
is given by
  \begin{eqnarray}
  U_r^{\dagger} | jm \rangle = \left[ 1-\delta(m,-j) \right] | j m-1 \rangle
  + \delta(m,-j) {\rm e}^{ - {\rm i} {\phi}_r} | j j \rangle
  \end{eqnarray}
  
We can check that the operator $H$ is Hermitean and the operator $U_r$ is 
unitary on the space $\varepsilon (j)$. Equation~(26) can be rewritten as
  \begin{eqnarray}
  \left( U_r \right)^{2j+1} = {\rm e}^{ {\rm i} {\phi}_r } I
  \end{eqnarray}  
which reflects the cyclic character of $U_r$ on $\varepsilon (j)$. 

Finally let us mention that, as far as the operators $H$, $U_r$ and 
$U_r^\dagger$ act on the space $\varepsilon (j)$, one can write
   \begin{eqnarray}
   H = \sum_{m=-j}^{j} {\sqrt{ (j+m)(j-m+1) }} |j m \rangle \langle j m |
   \\
   U_r = \sum_{m=-j}^{j-1} |j m+1 \rangle \langle j m| 
       + {\rm e}^{ + {\rm i} {\phi}_r} |j -j \rangle \langle j j|
   \\
   U_r^{\dagger} = \sum_{m=-j+1}^{j} | j m-1 \rangle \langle j m |
                 + {\rm e}^{ - {\rm i} {\phi}_r} | j j \rangle \langle j -j|
  \end{eqnarray}
where we have introduced {\em \`a la} Dirac projectors on $\varepsilon (j)$.

\subsection{The SU(2) generators}
We are now in a position to give a realization of the Lie algebra of the group 
SU(2) in terms of the generators of $A_1$ and $A_2$. Let us define the 
three operators
  \begin{equation}
  J_+ = H           U_r,  \quad  
  J_- = U_r^{\dagger} H
  \end{equation}  
and
  \begin{equation}
  J_z = {1 \over 2} \left( N_1-N_2 \right)
  \end{equation}
It is straightforward  to check that the action on the vector 
$ | jm \rangle $ of the operators defined by Eqs.~(40) and (41) 
is given by 
  \begin{eqnarray}
  J_+ |j m \rangle &=& {\sqrt{ (j - m)(j + m+1) }} |j m + 1 \rangle \\
  J_- |j m \rangle &=& {\sqrt{ (j + m)(j - m+1) }} |j m - 1 \rangle 
  \end{eqnarray}
and
  \begin{eqnarray}
  J_z   |j m \rangle = m |jm \rangle
  \end{eqnarray}
Consequently, we have the commutation relations 
  \begin{eqnarray}
  \left[ J_z,J_{+} \right] = + J_{+},  \quad 
  \left[ J_z,J_{-} \right] = - J_{-},  \quad 
  \left[ J_+,J_- \right] = 2J_z 
  \end{eqnarray}
which correspond to the Lie algebra of SU(2). 

We have here an unsual result 
for Lie algebras. In the context of deformations, we generally 
start from a Lie algebra, then deform it and finally find a 
realization in terms of deformed oscillator algebras. Here 
we started from two $q$-deformed oscillator algebras from which 
we derived the nondeformed Lie algebra su(2).  

\section{An alternative basis for the representation of SU(2)}
\subsection{An alternative to the $\{ J^2 , J_z\}$ scheme}
The decomposition (40) of the shift operators $J_+$ and $J_-$ in terms 
of $H$ and $U_r$ coincides with the polar
decomposition introduced in Ref.~[13] in a completely different 
way. This is easily seen by taking the matrix elements of
$U_r$ and $H$ in the $\{ J^2 , J_z \}$ quantization scheme  
and by comparing these elements to the ones of the operators 
$\Upsilon$ and $J_{\rm T}$ in Ref.~[13]. We are thus left with 
\begin{eqnarray}
H = J_{\rm T}
\end{eqnarray} 
and, by identifying the arbitrary phase $\varphi$ of Ref.~[13] 
with $\phi_r = 2 \pi j r = \pi (k-1)r$, we obtain that 
\begin{eqnarray}
U_r = \Upsilon
\end{eqnarray}  
so that Eq.~(40) corresponds to $J_+ = J_{\rm T} \Upsilon$ and 
$J_- = \Upsilon ^{\dagger} J_{\rm T}$.

It is immediate to check that the Casimir operator 
  \begin{eqnarray}
  J^2 = {1 \over 2} \left( J_+J_- +  
                           J_-J_+ \right) + J_z^2 
  \end{eqnarray}
of su(2) can be rewritten as
\begin{eqnarray}
J^2 = H^2 + J_z^2 - J_z = U_r ^{\dagger} H^2 U_r + J_z^2 + J_z
\end{eqnarray}
or 
\begin{eqnarray}
J^2 = \frac{1}{4} (N_1 + N_2) (N_1 + N_2 + 2) 
\end{eqnarray}
in terms of the generators $N_1$ and $N_2$ 
of $A_1$ and $A_2$, respectively. It is a simple matter of calculation
to prove that $J^2$ commutes with $U_r$ for any value of $r$. (Note that the commutator 
$[U_r, U_s]$ is different from zero for $r \ne s$.) Therefore, for $r$ fixed,
the commuting set $\{ J^2, U_r\}$ provides us with an alternative to the
familiar commuting set $\{ J^2, J_z \}$ of angular momentum theory. It is
to be observed that the operators 
$J^2$ and $U_r$ can be expressed as functions of the generators of $A_1$ 
and $A_2$ (see Eqs.~(19) and (50)).

\subsection{Eigenvalues and eigenvectors}
The next step is to determine the eigenvalues and eigenvectors of $U_r$. 
The eigenvalues and the common eigenvectors of the complete set of commuting 
operators $\{ J^2, U_r \}$ can be easily found. This leads to the following result.
The spectra of the operators $U_r$ and $J^2$ are given by
  \begin{eqnarray}
  U_r | j \alpha ; r \rangle &=& q^{-\alpha} 
      | j \alpha ; r \rangle \nonumber \\
  J^2 | j \alpha ; r \rangle &=& j(j+1) 
      | j \alpha ; r \rangle 
  \end{eqnarray}
where 
  \begin{equation}
  |j \alpha ; r \rangle = {1 \over {\sqrt{2j + 1}}} 
  \sum_{m = -j}^j
  q^{\alpha m} 
  |j m \rangle 
  \end{equation}
with the range of values 
  \begin{eqnarray}
  \alpha = - jr, - jr + 1, \cdots, -jr + 2j, \quad  2j \in {\bf N}, \quad r \in {\bf R}
  \end{eqnarray}
modulo $2j+1$. The parameter $q$ in Eqs.~(51) and (52) is 
  \begin{equation}
  q = \exp \left( {\rm i} \frac{2 \pi}{2 j + 1} \right) 
  \end{equation}
(cf.~Eq.~(8) with $k=2j+1$ for $k \in {\bf N} \setminus \{ 0,1 \}$ and 
$k \to \infty$ for $j=0$). 

The label $\mu$ used in Section 2 is here of the form
$\mu \equiv \alpha ; r$ with a fixed value of $r$. It is important to 
note that the label $\alpha$ in Eqs.~(51) and (52) goes, 
by step of 1, from $-jr$ to $-jr + 2j$; it is only for $r=1$ 
that  $\alpha$  goes, by step of 1, from $-j$ to $j$. 

The inter-basis expansion coefficients 
\begin{eqnarray}
\langle jm | j \alpha ; r \rangle = {1 \over \sqrt{2 j + 1}} q^{\alpha m} 
= {1 \over \sqrt{2 j + 1}} 
\exp \left( {\rm i} \frac{2 \pi}{2 j + 1} \alpha m \right)
\end{eqnarray}
(with 
$m      = -j , -j  + 1, \cdots,        j$ 
and 
$\alpha = -jr, -jr + 1, \cdots, -jr + 2j$) in Eq.~(52) define a unitary
transformation that allows to pass from the well-known 
orthonormal standard spherical basis 
\begin{eqnarray}
S   = \{ |j m \rangle : 2j \in {\bf N}, \ m = - j, - j + 1, \cdots, j \}
\end{eqnarray}
to the orthonormal nonstandard basis
\begin{eqnarray}
B_r = \{ |j \alpha ; r \rangle : 2j \in {\bf N}, \ 
\alpha = - jr, - jr + 1, \cdots, -jr + 2j \}
\end{eqnarray}
for the space 
\begin{eqnarray}
\varepsilon = \bigoplus_{j = 0, \frac{1}{2}, 1, \cdots} \varepsilon(j) 
\end{eqnarray}
where $\varepsilon(j)$ is a subspace of constant angular momentum $j$ (see Eq.~(32)).
For fixed $r$, the expansion coefficients satisfy the unitarity property
\begin{eqnarray}
\sum_{m=-j}^{j} 
\langle jm | j \alpha  ; r \rangle^* \>
\langle jm | j \alpha' ; r \rangle 
= \delta (\alpha' , \alpha)
\end{eqnarray}
and
\begin{eqnarray}
\sum_{\alpha=-jr}^{-jr+2j} 
\langle jm  | j \alpha ; r \rangle    \>
\langle jm' | j \alpha ; r \rangle^* 
= \delta (m' , m)
\end{eqnarray}
Then, the development 
  \begin{eqnarray}
  |j m \rangle = {1 \over {\sqrt{2j + 1}}} 
  \sum_{\alpha = -jr}^{-jr + 2j}
  q^{- m \alpha}                            \>
  |j \alpha ; r \rangle  
  \end{eqnarray}
with 
  \begin{eqnarray}
  m = - j, - j + 1, \cdots, j,  \quad  2j \in {\bf N}
  \end{eqnarray}
is the inverse of Eq.~(52) and makes it possible to pass from the
nonstandard basis $B_r$ to the standard basis $S$.

The representation theory 
of SU(2) can be transcribed in the $\{ J^2 , U_r \}$ scheme. In this 
scheme, the rotation matrix elements for the rotation $R$ of SO(3) 
assumes the form 
\begin{eqnarray}
{\cal D}_r^{(j)}(R)_{\alpha \alpha'} = \frac{1}{2j+1} 
\sum_{m  = -j}^{j}
\sum_{m' = -j}^{j}
q^{-\alpha m + \alpha' m'} \> 
{\cal D}^{(j)}(R)_{mm'}
\end{eqnarray} 
in terms of the standard matrix elements ${\cal D}^{(j)}(R)_{mm'}$. Then,
the behavior of the vector $|j \alpha ; r \rangle$ 
under an arbitrary rotation $R$ is given by 
\begin{eqnarray}
P_{R} |j \alpha ; r \rangle = \sum_{\alpha'} |j \alpha' ; r \rangle  \>
{\cal D}_r^{(j)}(R)_{\alpha' \alpha}
\end{eqnarray}
where $P_{R}$ stands for the operator 
associated with $R$. If $R$ is a rotation around the 
$z$-axis, Eq.~(64) takes a simple form. 
Indeed, if $R(\varphi)$ is a rotation of an angle 
\begin{eqnarray}
\varphi = p \frac{2 \pi}{2j+1} \quad {\rm with} \quad p = 0, 1, 2, \cdots, 2j
\end{eqnarray}
around the $z$-axis, we have
\begin{eqnarray}
P_{R(\varphi)} \> 
|j \alpha  ; r \rangle = 
|j \alpha' ; r \rangle
\end{eqnarray}
where
\begin{eqnarray}
\alpha' = \alpha - p, \quad {\rm mod}(2j+1)
\end{eqnarray}
Consequently, the set
$\{ |j \alpha ; r \rangle : \alpha = - jr, - jr + 1, \cdots, -jr + 2j \}$ 
spans a representation of dimension $2j+1$ of the cyclic subgroup 
$C_{2j+1}$ of SO(3). It can be seen that this representation is nothing but the
regular representation of $C_{2j+1}$. The nonstandard basis $B_r$ presents some
characteristics of a group-subgroup type basis in the sense that the set  
$\{ |j \alpha ; r \rangle : \alpha = - jr, - jr + 1, \cdots, -jr + 2j \}$  
carries a representation of a subgroup of SO(3). However, this representation 
is reducible except for $j=0$. Therefore, the label $\mu \equiv \alpha ; r$
does not correspond to some irreducible representation of a subgroup of 
SU(2) or ${\rm SO}(3) \equiv {\rm SU}(2)/Z_2$ so that the basis $B_r$ 
also exhibits some characteristics of a nongroup-subgroup type basis. 

The behavior of the vector $|j \alpha ; r \rangle$ under 
the time-reversal operator $K$ is given by 
\begin{eqnarray}
K |j \alpha ; r \rangle = \sum_{\alpha'} 
  \pmatrix{
  j     &j         \cr
  \alpha&\alpha'   \cr
  }_r
|j \alpha' ; r \rangle
\end{eqnarray}
where
\begin{eqnarray}
  \pmatrix{
  j     &j         \cr
  \alpha&\alpha'   \cr
  }_r = \frac{1}{2j+1}                        \>
  \sum_{m =-j}^j                              \>
  \sum_{m'=-j}^j                              \>
  q^{- \alpha m - \alpha' m'}                 \>
  \pmatrix{
  j     &j         \cr
  m     &m'        \cr
  }
\end{eqnarray} 
Here, the 2-$jm$ symbol (also called a 
1-$jm$ symbol for evident reasons) reads 
\begin{eqnarray}
  \pmatrix{
  j     &j         \cr
  m     &m'        \cr
  } = (-1)^{j+m} \delta(m' , -m)
\end{eqnarray}
and defines the metric tensor introduced by Wigner.$^6$ (The 
normalization chosen for the Wigner metric tensor is the one of 
Edmonds.$^{14}$)

The 2-$j\alpha$ metric tensor allows us to pass from a given 
irreducible representation matrix of SU(2)
to its complex conjugate. Indeed, we have
\begin{eqnarray}
  {\cal D}_r^{(j)}(R)_{\beta \beta'}^* = \sum_{\alpha \alpha'} 
  \pmatrix{
  j         &j              \cr
  \beta     &\alpha         \cr
  }_r^*                                                    \> 
  {\cal D}_r^{(j)}(R)_{\alpha \alpha'}                     \>
  \pmatrix{
  j          &j              \cr
  \beta'     &\alpha'        \cr
  }_r        
\end{eqnarray}
(the two $j$'s in the 2-$j\alpha$ metric tensor are identical 
because the irreducible representation class ($j$)  of  SU(2) 
is identical to its complex conjugate).  

For any value of $r$, the basis $B_r$ is an alternative to the spherical basis
$S$ of the space $\varepsilon$. Two bases $B_r$ and $B_s$ with $r \not=s$ are 
thus two equally admissible orthonormal bases for $\varepsilon$. The 
vectors of the bases $B_r$ and $B_s$ are common eigenvectors of 
$\{ J^2 , U_r \}$ and 
$\{ J^2 , U_s \}$, respectively. The overlap between the
bases $B_r$ and $B_s$ is controlled by
\begin{eqnarray}
\langle j' \alpha ; r | j \beta ; s \rangle = \delta (j' , j) \> \frac{1}{2j+1} \> 
\frac{\sin (\alpha - \beta) \pi}
     {\sin (\alpha - \beta) \frac{\pi}{2j+1}}
\end{eqnarray}
with $\alpha = - jr, - jr + 1, \cdots, -jr + 2j$ 
and  $\beta  = - js, - js + 1, \cdots, -js + 2j$.

\subsection{Some examples}
As an illustration, we continue with some examples concerning the subspaces
$\varepsilon(\frac{1}{2})$ and $\varepsilon(1)$.

\subsubsection{The case $j = \frac{1}{2}$}
For $r=1$, Eq.~(52) gives
 \begin{eqnarray}
 |\frac{1}{2} -\frac{1}{2} ; 1 \rangle &=& \frac{1}{\sqrt{2}} \left( 
 \rho      |\frac{1}{2} -\frac{1}{2} \rangle + 
 \rho^{-1} |\frac{1}{2} +\frac{1}{2} \rangle \right)                 \nonumber \\ 
 |\frac{1}{2} +\frac{1}{2} ; 1 \rangle &=& \frac{1}{\sqrt{2}} \left( 
 \rho^{-1} |\frac{1}{2} -\frac{1}{2} \rangle + 
 \rho      |\frac{1}{2} +\frac{1}{2} \rangle \right)
 \end{eqnarray}
where $\rho = {\rm e}^{{\rm i} \frac{\pi}{4}}$. For $r=0$, we have
 \begin{eqnarray}
 |\frac{1}{2} 0 ; 0 \rangle &=& \frac{1}{\sqrt{2}} \left( 
 |\frac{1}{2} -\frac{1}{2} \rangle + 
 |\frac{1}{2} +\frac{1}{2} \rangle \right)                           \nonumber \\ 
 |\frac{1}{2} 1 ; 0 \rangle &=& \frac{1}{\sqrt{2}} \left( 
 \rho^{-2} |\frac{1}{2} -\frac{1}{2} \rangle + 
 \rho^2    |\frac{1}{2} +\frac{1}{2} \rangle \right)
 \end{eqnarray}

\subsubsection{The case $j = 1$}
By putting 
$\omega = {\rm e}^{{\rm i} \frac{2 \pi}{3}}$, we obtain
 \begin{eqnarray}
 |1 -1 ; 1 \rangle &=& \frac{1}{\sqrt{3}} \left( 
 \omega |1 -1 \rangle + |1 0 \rangle + \omega^{-1} |1  +1 \rangle \right) \nonumber \\ 
 |1  0 ; 1 \rangle &=& \frac{1}{\sqrt{3}} \left( 
 |1 -1 \rangle + |1 0 \rangle + |1  +1 \rangle \right)                    \nonumber \\ 
 |1 +1 ; 1 \rangle &=& \frac{1}{\sqrt{3}} \left( 
 \omega^{-1} |1 -1 \rangle + |1 0 \rangle + \omega |1  +1 \rangle \right) 
 \end{eqnarray}
for $r=1$ and
\begin{eqnarray}
 |1  0 ; 0 \rangle &=& \frac{1}{\sqrt{3}} \left( 
 |1 -1 \rangle + |1 0 \rangle + |1  +1 \rangle \right)                    \nonumber \\ 
 |1  1 ; 0 \rangle &=& \frac{1}{\sqrt{3}} \left( 
 \omega^{-1} |1 -1 \rangle + |1 0 \rangle + \omega |1  +1 \rangle \right) \nonumber \\ 
 |1  2 ; 0 \rangle &=& \frac{1}{\sqrt{3}} \left( 
 \omega |1 -1 \rangle + |1 0 \rangle + \omega^{-1} |1  +1 \rangle \right) 
 \end{eqnarray}
for $r=0$. 
                           
We thus foresee that it is quite possible to achieve the construction of 
the WRa of the group
SU(2) in the $\{ J^2, U_r \}$ scheme. This furnishes an alternative to the
WRa of SU(2) in the SU(2) $\supset$ U(1) basis corresponding 
to the $\{ J^2, J_z \}$ scheme. 

\section{A new approach to the Wigner-Racah algebra of SU(2)}
In this section, we give the basic
ingredients for the WRa of SU(2) in the $\{ J^2, U_r \}$ scheme. 
The Clebsch-Gordan coefficients (CGc's) or coupling coefficients 
adapted to the $\{ J^2, U_r \}$ scheme
are defined from the SU(2) $\supset$ U(1) CGc's adapted to the $\{ J^2, J_z \}$ 
scheme. The adaptation to the $\{ J^2, U_r \}$ scheme afforded by Eq.~(52) is
transferred to SU(2) irreducible tensor operators. This yields the
Wigner-Eckart theorem in the $\{ J^2, U_r \}$ scheme.  

\subsection{Coupling coefficients in the $\{ J^2, U_r \}$ scheme}
When passing from the $\{ J^2, J_z \}$ scheme to the $\{ J^2, U_r \}$ scheme, 
the CGc's $( j_1 j_2 m_1 m_2 | j_3 m_3 )$ are replaced by the coefficients 
  \begin{eqnarray}
  \left( j_1 j_2 \alpha_1 \alpha_2 |j_3 \alpha_3 \right)_r = 
  {1 \over \sqrt{(2j_1 + 1) 
                 (2j_2 + 1) 
		 (2j_3 + 1)}}                    \nonumber          \\ 
  \sum_{m_1=-j_1}^{j_1} 
  \sum_{m_2=-j_2}^{j_2} 
  \sum_{m_3=-j_3}^{j_3} 
         q_1^{- \alpha_1 m_1}      \>
         q_2^{- \alpha_2 m_2}      \>
	 q_3^{  \alpha_3 m_3}      \>
   ( j_1 j_2 m_1 m_2 | j_3 m_3 ) 
  \end{eqnarray}
where the $q_a$'s are given in terms of $j_a$ by
       \begin{eqnarray}
q_a = \exp \left( {\rm i} {2 \pi \over 2 j_a + 1 } \right), \quad a = 1,2,3
       \end{eqnarray} 
(cf. Eq.~(54)).
      
      The new CGc's  
$( j_1 j_2 \alpha_1 \alpha_2 | j \alpha )_r$ in the $\{ J^2, U_r \}$ scheme 
are simple linear combinations of the SU(2) $\supset$ U(1) CGc's. The 
symmetry properties of the coupling coefficients 
$( j_1 j_2 \alpha_1 \alpha_2 | j \alpha )_r$
cannot be expressed in a simple way
(except the symmetry under the interchange 
$j_1 \alpha_1 \leftrightarrow 
 j_2 \alpha_2$). 
Let us introduce the $f_r$ symbol via 
  \begin{eqnarray}
  f_r\pmatrix{
  j_1     &j_2     &j_3     \cr
  \alpha_1&\alpha_2&\alpha_3\cr
  }
  =      (-1)^{2j_3} {1 \over {\sqrt{2j_1+1}}}
  \left( j_2 j_3 \alpha_2 \alpha_3 | j_1 \alpha_1 \right)_r^*
  \end{eqnarray}
Its value is multiplied by the factor 
$(-1)^{j_1 + j_2 + j_3}$ when its two last columns
are interchanged. However, the interchange of two other columns cannot be
described by a simple symmetry property. Nevertheless, the $f_r$ symbol is of
central importance for the calculation of matrix 
elements of irreducible tensor operators via 
the Wigner-Eckart theorem in the $\{ J^2 , U_r \}$ 
scheme (see Eq.~(106) below).  

Following Ref.~[9], we define a more
symmetrical symbol, namely the $\overline{f_r}$ symbol, through 
  \begin{eqnarray}
  \overline{f_r} \pmatrix{
  j_1     &j_2     &j_3     \cr
  \alpha_1&\alpha_2&\alpha_3\cr
  } = 
  {1 \over {\sqrt{(2j_1 + 1) (2j_2 + 1) (2j_3 + 1)} } } \nonumber \\
  \sum_{m_1 = -j_1}^{j_1} 
  \sum_{m_2 = -j_2}^{j_2} 
  \sum_{m_3 = -j_3}^{j_3}  
         q_1^{ - \alpha_1 m_1 } \>
         q_2^{ - \alpha_2 m_2 } \>
         q_3^{ - \alpha_3 m_3 } 
  \pmatrix{
  j_1&j_2&j_3\cr
  m_1&m_2&m_3\cr
  }
  \end{eqnarray}
The 3-$jm$ symbol on the right-hand side of Eq.~(80) is an ordinary 
Wigner symbol for the SU(2) group in the SU(2)~$\supset$~U(1) basis.
It is possible to 
pass from the $f_r$ symbol to the $\overline{f_r}$ symbol and vice versa by means
of the metric tensor introduced in Section 4. Indeed, we can check that 
\begin{eqnarray}
  \overline{f_r} \pmatrix{
  j_1     &j_2     &j_3     \cr
  \alpha_1&\alpha_2&\alpha_3\cr
                   } = \sum_{\alpha'_3} \>
  \pmatrix{
  j_3     &   j_3      \cr
  \alpha_3&   \alpha'_3\cr
        }_r                             \>
  f_r \pmatrix{
  j_3      &j_2     &j_1     \cr
  \alpha_3'&\alpha_2&\alpha_1\cr
               }^*
\end{eqnarray}
or alternatively
\begin{eqnarray}
  f_r \pmatrix{
  j_1     &j_2     &j_3     \cr
  \alpha_1&\alpha_2&\alpha_3\cr
                   } = \sum_{\alpha'_1}    \>
  \pmatrix{
  j_1      &   j_1      \cr
  \alpha_1'&   \alpha_1 \cr
        }_r                                \>
  \overline{f_r} \pmatrix{
  j_1      &j_3     &j_2     \cr
  \alpha_1'&\alpha_3&\alpha_2\cr
               }^*
\end{eqnarray}
The $\overline{f_r}$ symbol is more symmetrical than the $f_r$ symbol.
The $\overline{f_r}$ symbol exhibits the same symmetry properties under 
permutations of its columns as the 3-$jm$ Wigner symbol: Its value 
is multiplied by $(-1)^{j_1 + j_2 + j_3}$ under an odd permutation 
and does not change under an even permutation. In other words, we have
\begin{eqnarray}
  \overline{f_r} \pmatrix{
  j_1     &j_2     &j_3     \cr
  \alpha_1&\alpha_2&\alpha_3\cr
                   } =   \varepsilon_{abc} \>
  \overline{f_r} \pmatrix{
  j_a      &j_b      &j_c      \cr
  \alpha_a &\alpha_b &\alpha_c \cr
               }
\end{eqnarray}
where $\varepsilon_{abc} = 1$ or $(-1)^{j_1 + j_2 + j_3}$ according to 
whether $abc$ corresponds to an even or odd permutation of $123$.

The orthogonality properties of the highly symmetrical 
$\overline{f_r}$ symbol easily follow from  the corresponding properties of the 
3-$jm$ Wigner symbol. Thus, we have 
  \begin{eqnarray}
  \sum_{j_3 \alpha_3}                    \> 
  (2j_3 +1)                              \>
  \overline{f_r} \pmatrix{
  j_1     &j_2     &j_3     \cr
  \alpha_1&\alpha_2&\alpha_3\cr
  }^*                                    \>
  \overline{f_r} \pmatrix{
  j_1      &j_2      &j_3     \cr
  \alpha_1'&\alpha_2'&\alpha_3\cr
  }                                         \nonumber \\
     = \delta (\alpha_1' , \alpha_1)
       \delta (\alpha_2' , \alpha_2)
  \end{eqnarray}  
and
  \begin{eqnarray}
  \sum_{\alpha_1 \alpha_2}           \>
  \overline{f_r} \pmatrix{
  j_1     &j_2     &j_3     \cr
  \alpha_1&\alpha_2&\alpha_3\cr
  }                                  \>
  \overline{f_r} \pmatrix{
  j_1     &j_2     &j_3'     \cr
  \alpha_1&\alpha_2&\alpha_3'\cr
  }^*                                       \nonumber       \\ = 
  {1 \over 2 j_3 + 1} \> \Delta ( 0 | j_1 \otimes j_2 \otimes j_3 ) \>
  \delta (j_3' , j_3) \> \delta (\alpha_3' , \alpha_3)
  \end{eqnarray}  
where $\Delta ( 0 | j_1 \otimes j_2 \otimes j_3 ) =1$ or 0 according 
to whether the Kronecker product $(j_1) \otimes (j_2) \otimes (j_3)$ 
contains or does not contain 
the identity irreducible representation class (0) of SU(2). 
Note that the real number $r$ is the same for all the $\overline{f_r}$ symbols 
occurring in Eqs.~(84) and (85). 

The values of the SU(2) CGc's in the $\{ J^2, U_r \}$ scheme as well
as of the $f_r$ and $\overline{f_r}$ coefficients are not necessarily real 
numbers. For instance, we have the following property under complex conjugation
\begin{eqnarray}
  \overline{f_r} \pmatrix{
  j_1      &j_2      &j_3      \cr
  \alpha_1'&\alpha_2'&\alpha_3'\cr
              }^* = \sum_{\alpha_1 \alpha_2 \alpha_3} \>
  \pmatrix{
  j_1      &   j_1     \cr
  \alpha_1'&   \alpha_1\cr
        }_r^*                                         \>	
  \pmatrix{
  j_2      &   j_2     \cr
  \alpha_2'&   \alpha_2\cr
        }_r^*                                         \> 	
  \pmatrix{
  j_3      &   j_3     \cr
  \alpha_3'&   \alpha_3\cr
        }_r^* 		                    \nonumber     \\
  \overline{f_r} \pmatrix{
  j_1     &j_2     &j_3     \cr
  \alpha_1&\alpha_2&\alpha_3\cr
               }
\end{eqnarray}
Then, the behavior of the 
$\overline{f_r}$ symbol under complex conjugation is completely different from 
the one of the ordinary 3-$jm$ Wigner symbol. In this respect, we have 
  \begin{eqnarray}
  \overline{f_r} \pmatrix{
  j_1     &j_2     &j_3     \cr
  \alpha_1&\alpha_2&\alpha_3\cr
  }^* =  (-1)^{j_1 + j_2 + j_3}  \>
  \overline{f_r} \pmatrix{
  j_1     &j_2     &j_3     \cr
  \alpha_1&\alpha_2&\alpha_3\cr
  }
  \end{eqnarray}
Hence, the value of the $\overline{f_r}$ coefficient 
is real if $j_1 + j_2 + j_3$ is 
even and pure imaginary if  $j_1 + j_2 + j_3$ is odd. 

It is to be noted that the 2-$j\alpha$ symbol introduced in Section 4
is a particular case of the $\overline{f_r}$ symbol since we have
\begin{eqnarray}
  \pmatrix{
  j     &j       \cr
  \alpha&\alpha' \cr
              }_r   =   \sqrt{2j+1} \> 
  \overline{f_r} \pmatrix{
  j     &0     &j      \cr
  \alpha&0     &\alpha'\cr
  }  
\end{eqnarray}
Consequently, the orthogonality property 
\begin{eqnarray}
  \sum_{\alpha}                   \>
  \pmatrix{
  j     &j        \cr
  \alpha&\beta    \cr
              }_r                 \>
  \pmatrix{	      
  j       &j         \cr
  \alpha  &\beta'    \cr
              }_r^*	 = \delta( \beta' , \beta )     
\end{eqnarray}
and the symmetry property
\begin{eqnarray}
  \pmatrix{
  j      &j       \cr
  \alpha'&\alpha  \cr
              }_r =   (-1)^{2j} \>
  \pmatrix{
  j       &j        \cr
  \alpha  &\alpha'  \cr
              }_r	      
\end{eqnarray}
follow from the corresponding properties of the $\overline{f_r}$ symbol. 

The case $r=1$ deserves a special attention. In that case, we have specific 
relations because the label $\alpha$ may be $0$ for $j$ integer. For example,  
the value of 
\begin{eqnarray}
  \overline{f_r} \pmatrix{
  j_1     &j_2     &j_3     \cr
  0       &0       &0       \cr
  } = 
  {1 \over {\sqrt{(2j_1 + 1) (2j_2 + 1) (2j_3 + 1)} } } \nonumber \\
  \sum_{m_1 = -j_1}^{j_1} 
  \sum_{m_2 = -j_2}^{j_2} 
  \sum_{m_3 = -j_3}^{j_3}  
  \pmatrix{
  j_1&j_2&j_3\cr
  m_1&m_2&m_3\cr
  }
  \end{eqnarray}
is equal to $0$ if $j_1+j_2+j_3$ is odd.

\subsection{Recoupling coefficients in the $\{ J^2, U_r \}$ scheme}
The recoupling coefficients of the SU(2) group are rotational 
invariants.$^{14}$ Therefore, they can be
expressed in terms of coupling coefficients of SU(2) in the  
$\{ J^2 , U_r \}$ scheme. For example, the 9-$j$ symbol can be expressed 
in terms of $\overline{f_r}$ symbols by replacing, in its decomposition in terms of
3-$jm$ symbols, the 3-$jm$ symbols by $\overline{f_r}$ symbols. On the other hand,
the decomposition of  the 6-$j$ symbol   in terms of  
$\overline{f_r}$ symbols requires the introduction of six
metric tensors corresponding to the six arguments of the 6-$j$ symbol. 
These matters shall be developed by following the approach initiated in 
Ref.~[9].

We start with the case of the 6-$j$ symbol. Relations involving the 
6-$j$ Wigner symbol (or $\overline{W}$ Fano and Racah coefficient$^7$) 
and $\overline{f_r}$ symbols, with four 
    $\overline{f_r}$ symbols, can be easily derived. First, the 
6-$j$ symbol can be expressed as 
\begin{eqnarray}
\overline{W} \pmatrix{
  j_1     &j_2     &j_3     \cr
  j_4     &j_5     &j_6     \cr
               } = \sum_{{\rm all} \ \alpha'}
	           \sum_{{\rm all} \ \alpha}             
  \pmatrix{
  j_1      &j_1      \cr
  \alpha_1 &\alpha_1'\cr
          }_r^* 	
  \pmatrix{
  j_2      &j_2      \cr
  \alpha_2 &\alpha_2'\cr
          }_r^* 
  \pmatrix{
  j_3      &j_3      \cr
  \alpha_3 &\alpha_3'\cr
          }_r^*                                      \nonumber \\
  \pmatrix{
  j_4      &j_4      \cr
  \alpha_4 &\alpha_4'\cr
          }_r^* 
  \pmatrix{
  j_5      &j_5      \cr
  \alpha_5 &\alpha_5'\cr
          }_r^* 
  \pmatrix{
  j_6      &j_6      \cr
  \alpha_6 &\alpha_6'\cr
          }_r^* 		                     \nonumber \\
  \overline{f_r} \pmatrix{
  j_1     &j_2     &j_3     \cr
  \alpha_1&\alpha_2&\alpha_3\cr
                   }	      
  \overline{f_r} \pmatrix{
  j_1      &j_5     &j_6      \cr
  \alpha_1'&\alpha_5&\alpha_6'\cr
                   }	                             \nonumber \\
  \overline{f_r} \pmatrix{
  j_4      &j_2      &j_6     \cr
  \alpha_4'&\alpha_2'&\alpha_6\cr
                   }		   	      
  \overline{f_r} \pmatrix{
  j_4     &j_5      &j_3      \cr
  \alpha_4&\alpha_5'&\alpha_3'\cr
                   }
\end{eqnarray}
which involves 0$+$4 $\overline{f_r}$ symbols (no $\overline{f_r}$ symbol on the 
left-hand side and four on the right-hand side). With the help of Eq.~(86), 
Eq.~(92) can be rewritten as 
\begin{eqnarray}
\overline{W} \pmatrix{
  j_1     &j_2     &j_3     \cr
  j_4     &j_5     &j_6     \cr
               } = 
	           \sum_{\alpha_4' \alpha_5' \alpha_6'}
		   \sum_{{\rm all} \ \alpha} 
  \pmatrix{
  j_4      &j_4      \cr
  \alpha_4 &\alpha_4'\cr
          }_r^* 
  \pmatrix{
  j_5      &j_5      \cr
  \alpha_5 &\alpha_5'\cr
          }_r^* 
  \pmatrix{
  j_6      &j_6      \cr
  \alpha_6 &\alpha_6'\cr
          }_r^* 		           \nonumber \\
  \overline{f_r} \pmatrix{
  j_1     &j_2     &j_3     \cr
  \alpha_1&\alpha_2&\alpha_3\cr
                   }^*	      
  \overline{f_r} \pmatrix{
  j_1      &j_5     &j_6      \cr
  \alpha_1 &\alpha_5&\alpha_6'\cr
                   }	                   \nonumber \\     
  \overline{f_r} \pmatrix{
  j_4      &j_2      &j_6     \cr
  \alpha_4'&\alpha_2 &\alpha_6\cr
                   }		   	      
  \overline{f_r} \pmatrix{
  j_4     &j_5      &j_3      \cr
  \alpha_4&\alpha_5'&\alpha_3 \cr
                   }
\end{eqnarray}
An expression involving 1$+$3 $\overline{f_r}$ symbols is
\begin{eqnarray}
  \overline{f_r} \pmatrix{
  j_1     &j_2     &j_3     \cr
  \alpha_1&\alpha_2&\alpha_3\cr
                   }	      
\overline{W} \pmatrix{
  j_1     &j_2     &j_3     \cr
  j_4     &j_5     &j_6     \cr
               } = 
	       \Delta ( 0 | j_1 \otimes j_2 \otimes j_3 )
	       \sum_{\alpha_4' \alpha_5' \alpha_6'} 
	       \sum_{\alpha_4  \alpha_5  \alpha_6 }           \nonumber    \\
  \pmatrix{
  j_4      &j_4      \cr
  \alpha_4 &\alpha_4'\cr
          }_r^* 
  \pmatrix{
  j_5      &j_5      \cr
  \alpha_5 &\alpha_5'\cr
          }_r^* 
  \pmatrix{
  j_6      &j_6      \cr
  \alpha_6 &\alpha_6'\cr
          }_r^* 		                              \nonumber   \\
  \overline{f_r} \pmatrix{
  j_1      &j_5     &j_6      \cr
  \alpha_1 &\alpha_5&\alpha_6'\cr
                   }	      
  \overline{f_r} \pmatrix{
  j_4      &j_2      &j_6     \cr
  \alpha_4'&\alpha_2 &\alpha_6\cr
                   }		   	      
  \overline{f_r} \pmatrix{
  j_4     &j_5      &j_3      \cr
  \alpha_4&\alpha_5'&\alpha_3 \cr
                   }
\end{eqnarray}
We also have a 2$+$2 relationship 
\begin{eqnarray}
\sum_{j_3 \alpha_3} (2j_3+1) 
  \overline{f_r} \pmatrix{
  j_1     &j_2     &j_3     \cr
  \alpha_1&\alpha_2&\alpha_3\cr
                   }
  \overline{f_r} \pmatrix{
  j_4     &j_5      &j_3      \cr
  \alpha_4&\alpha_5 &\alpha_3 \cr
                   }^*		   	      
\overline{W} \pmatrix{
  j_1     &j_2     &j_3     \cr
  j_4     &j_5     &j_6     \cr
               }                                       \nonumber \\
	       = 
	       \sum_{\alpha_4' \alpha_5' \alpha_6'}
	       \sum_{\alpha_6 }                             
  \pmatrix{
  j_4       &j_4      \cr
  \alpha_4  &\alpha_4'\cr
          }_r^*
  \pmatrix{
  j_5       &j_5      \cr
  \alpha_5' &\alpha_5\cr
          }_r^* 
  \pmatrix{
  j_6      &j_6      \cr
  \alpha_6 &\alpha_6'\cr
          }_r^* 		                       \nonumber \\
  \overline{f_r} \pmatrix{
  j_1      &j_5      &j_6      \cr
  \alpha_1 &\alpha_5'&\alpha_6'\cr
                   }	      
  \overline{f_r} \pmatrix{
  j_4      &j_2      &j_6     \cr
  \alpha_4'&\alpha_2 &\alpha_6\cr
                   }		   	      
\end{eqnarray}
and a 3$+$1 relationship 
\begin{eqnarray}
\sum_{j_3 \alpha_3}   
	       \sum_{\alpha_4' \alpha_5' \alpha_6'}
	       \sum_{\alpha_1  \alpha_5           }     
	       (2j_3+1)                        
  \pmatrix{
  j_4       &j_4      \cr
  \alpha_4'  &\alpha_4\cr
          }_r
  \pmatrix{
  j_5       &j_5      \cr
  \alpha_5  &\alpha_5'\cr
          }_r 
  \pmatrix{
  j_6      &j_6      \cr
  \alpha_6 &\alpha_6'\cr
          }_r 	                                \nonumber \\	         
  \overline{f_r} \pmatrix{
  j_1     &j_2     &j_3     \cr
  \alpha_1&\alpha_2&\alpha_3\cr
                   } 
  \overline{f_r} \pmatrix{
  j_1      &j_5      &j_6      \cr
  \alpha_1 &\alpha_5 &\alpha_6'\cr
                   }^*	      		   	   
  \overline{f_r} \pmatrix{
  j_4       &j_5       &j_3      \cr
  \alpha_4' &\alpha_5' &\alpha_3 \cr
                   }^*		   		\nonumber \\   	      
\overline{W} \pmatrix{
  j_1     &j_2     &j_3     \cr
  j_4     &j_5     &j_6     \cr                    
               }  
= \frac{1}{2j_6 + 1} \> \Delta ( 0 | j_1 \otimes j_5 \otimes j_6 ) \>
  \overline{f_r} \pmatrix{
  j_4      &j_2      &j_6     \cr
  \alpha_4 &\alpha_2 &\alpha_6\cr
                   }		   	      
\end{eqnarray}
By using the orthonormality of the $\overline{f_r}$ symbol in conjunction 
with Eq.~(96), we would obtain a 4$+$0 relationship 
which turns out to be the well-known orthonormality 
relation$^7$ for the $\overline{W}$ coefficient.

We continue with the 9-$j$ 
Wigner symbol (or $X$ Fano and Racah coefficient$^7$). Relations 
involving six $\overline{f_r}$ symbols and one 9-$j$ symbol can be 
obtained in a straightforward way. First, we have the 
very symmetrical expression of the type 0$+$6
\begin{eqnarray}	         
  X \pmatrix{
  j_{11}     &j_{12}     &j_{13}     \cr
  j_{21}     &j_{22}     &j_{23}     \cr
  j_{31}     &j_{32}     &j_{33}     \cr
                   } = \sum_{{\rm all} \ \alpha} 
\overline{f_r} \pmatrix{
  j_{11}       &j_{21}       &j_{31}      \cr
  \alpha_{11}  &\alpha_{21}  &\alpha_{31} \cr
                   }	      		   	   
  \overline{f_r} \pmatrix{
  j_{12}       &j_{22}       &j_{32}      \cr
  \alpha_{12}  &\alpha_{22}  &\alpha_{32} \cr
                   }	                       \nonumber \\   		   	      
  \overline{f_r} \pmatrix{
  j_{13}       &j_{23}       &j_{33}      \cr
  \alpha_{13}  &\alpha_{23}  &\alpha_{33} \cr
                   }		                  		   	   
  \overline{f_r} \pmatrix{
  j_{11}       &j_{12}       &j_{13}      \cr
  \alpha_{11}  &\alpha_{12}  &\alpha_{13} \cr
                   }^*                         \nonumber \\	      		   	   
  \overline{f_r} \pmatrix{
  j_{21}       &j_{22}       &j_{23}      \cr
  \alpha_{21}  &\alpha_{22}  &\alpha_{23} \cr
                   }^*		   		   	      
  \overline{f_r} \pmatrix{
  j_{31}       &j_{32}       &j_{33}      \cr
  \alpha_{31}  &\alpha_{32}  &\alpha_{33} \cr
                   }^*
\end{eqnarray}
Other relations with six $\overline{f_r}$ symbols can be derived by 
combining Eq.~(97) 
and the orthonormality relations of the $\overline{f_r}$ 
symbols. For instance, we have the relation of the type 1$+$5
\begin{eqnarray}	         
  \overline{f_r} \pmatrix{
  j_{31}       &j_{32}       &j_{33}      \cr
  \alpha_{31}  &\alpha_{32}  &\alpha_{33} \cr
                   }
  X \pmatrix{
  j_{11}     &j_{12}     &j_{13}     \cr
  j_{21}     &j_{22}     &j_{23}     \cr
  j_{31}     &j_{32}     &j_{33}     \cr
                   } = \Delta ( 0 | j_{31} \otimes j_{32} \otimes j_{33} ) 
		   \nonumber \\
		   \sum_{\alpha_{11}  \alpha_{12}  \alpha_{13}}
		   \sum_{\alpha_{21}  \alpha_{22}  \alpha_{23}}             
\overline{f_r} \pmatrix{
  j_{11}       &j_{21}       &j_{31}      \cr
  \alpha_{11}  &\alpha_{21}  &\alpha_{31} \cr
                   }	                                                   
		   \nonumber \\   		   	   
  \overline{f_r} \pmatrix{
  j_{12}       &j_{22}       &j_{32}      \cr
  \alpha_{12}  &\alpha_{22}  &\alpha_{32} \cr
                   }	   		   	      
  \overline{f_r} \pmatrix{
  j_{13}       &j_{23}       &j_{33}      \cr
  \alpha_{13}  &\alpha_{23}  &\alpha_{33} \cr
                   }		                                           
		   \nonumber \\		   	   
  \overline{f_r} \pmatrix{
  j_{11}       &j_{12}       &j_{13}      \cr
  \alpha_{11}  &\alpha_{12}  &\alpha_{13} \cr
                   }^*	      		   	   
  \overline{f_r} \pmatrix{
  j_{21}       &j_{22}       &j_{23}      \cr
  \alpha_{21}  &\alpha_{22}  &\alpha_{23} \cr
                   }^*		   		   	      
\end{eqnarray}
and the relation of the type 2$+$4
\begin{eqnarray}
\sum_{ j_{31} \alpha_{31} } (2j_{31} + 1)
  \overline{f_r} \pmatrix{
  j_{11}       &j_{21}       &j_{31}      \cr
  \alpha_{11}  &\alpha_{21}  &\alpha_{31} \cr
                   }^*	      		   		         
  \overline{f_r} \pmatrix{
  j_{31}       &j_{32}       &j_{33}      \cr
  \alpha_{31}  &\alpha_{32}  &\alpha_{33} \cr
                   }
		   \nonumber \\
  X \pmatrix{
  j_{11}     &j_{12}     &j_{13}     \cr
  j_{21}     &j_{22}     &j_{23}     \cr
  j_{31}     &j_{32}     &j_{33}     \cr
                   }                                               
		 = \sum_{ \alpha_{12}  \alpha_{13} }
		   \sum_{ \alpha_{22}  \alpha_{23} } 
		   \nonumber \\            
  \overline{f_r} \pmatrix{
  j_{12}       &j_{22}       &j_{32}      \cr
  \alpha_{12}  &\alpha_{22}  &\alpha_{32} \cr
                   }
  \overline{f_r} \pmatrix{
  j_{13}       &j_{23}       &j_{33}      \cr
  \alpha_{13}  &\alpha_{23}  &\alpha_{33} \cr
                   }		                                 
		   \nonumber \\		   	   
  \overline{f_r} \pmatrix{
  j_{11}       &j_{12}       &j_{13}      \cr
  \alpha_{11}  &\alpha_{12}  &\alpha_{13} \cr
                   }^*	      		   	   
  \overline{f_r} \pmatrix{
  j_{21}       &j_{22}       &j_{23}      \cr
  \alpha_{21}  &\alpha_{22}  &\alpha_{23} \cr
                   }^*		   		   	      
\end{eqnarray}

Relations involving coupling and recoupling 
coefficients are of considerable interest for
the calculation of matrix elements. In particular, 
$\overline{W}$ and $X$ coefficients occur in matrix 
elements of scalar product and tensor product of 
two irreducible tensor operators. 

\subsection{Wigner-Eckart theorem in the $\{ J^2, U_r \}$ scheme}
\subsubsection{Irreducible tensor operators}
From the spherical components $T^{(k)}_m$ (with $m = -k, -k+1, \cdots, k$) 
of an SU(2) irreducible tensor operator ${\bf T}^{(k)}$, we define the 
$2 k + 1$ components 
  \begin{eqnarray}
  T^{(k)}_{\alpha ; r} = {1 \over {\sqrt{2k+1}}} \sum_{m=-k}^k q^{\alpha m} \>
  T^{(k)}_m 
  \end{eqnarray}
with
  \begin{eqnarray}
\alpha = -kr, -kr + 1, \cdots, -kr + 2k,  \quad  2k \in {\bf N}
  \end{eqnarray}
where $r$ is fixed in ${\bf R}$. The behavior of $T^{(k)}_{\alpha ; r}$ 
under a rotation $R$ is described by 
\begin{eqnarray}
P_{R} \> T^{(k)}_{\alpha ; r} \> P_R^{-1} = \sum_{\alpha'} T^{(k)}_{\alpha' ; r}  \>
{\cal D}_r^{(j)}(R)_{\alpha' \alpha}
\end{eqnarray} 

Following Racah,$^7$ given two SU(2) irreducible tensor operators 
${\bf T}^{(k_1)}$ and 
${\bf U}^{(k_2)}$, we can define the tensor product 
$\{ {\bf T}^{(k_1)}{\bf U}^{(k_2)} \}^{(k)}$ of components 
\begin{eqnarray}
\{ {\bf T}^{(k_1)}{\bf U}^{(k_2)} \}^{(k)}_{\alpha ; r} = 
\sum_{\alpha_1 \alpha_2} (k_1 k_2 \alpha_1 \alpha_2 | k \alpha)_r \>
{T}^{(k_1)}_{\alpha_1 ; r}  \>
{U}^{(k_2)}_{\alpha_2 ; r}
\end{eqnarray}
As a particular case, we get the scalar product 
\begin{eqnarray}
\left( {\bf T}^{(k)} \cdot {\bf U}^{(k)}\right) = (-1)^{k} \> \sqrt{2k+1} \>
\{ {\bf T}^{(k)}{\bf U}^{(k)} \}^{(0)}_{0 ; r}
\end{eqnarray}
More specifically, we have
\begin{eqnarray}
\left( {\bf T}^{(k)} \cdot {\bf U}^{(k)}\right) = (-1)^{-k} \sum_{\alpha \alpha'}
\pmatrix{
  k      &k         \cr
  \alpha &\alpha'   \cr
  }_r                     \> 
  {T}^{(k)}_{\alpha  ; r} \> 
  {U}^{(k)}_{\alpha' ; r}
\end{eqnarray}
which can be identified with the scalar 
product introduced by Racah.$^7$

\subsubsection{Matrix elements of tensor operators}  
In the $\{ J^2, U_r \}$ scheme, the Wigner-Eckart theorem reads
 \begin{eqnarray}
 \langle \tau_1 j_1 \alpha_1 ; r | T^{(k)}_{\alpha ; r} | 
         \tau_2 j_2 \alpha_2 ; r \rangle = 
         \left( 
         \tau_1 j_1             || T^{(k)}               || 
         \tau_2 j_2 \right) \>
 f_r\pmatrix{
 j_1     &j_2     &k     \cr
 \alpha_1&\alpha_2&\alpha\cr
 }
 \end{eqnarray}
where $\left( \tau_1 j_1 || T^{(k)} || \tau_2 j_2 \right)$ 
denotes an ordinary reduced matrix
element. Such a 
reduced matrix element is clearly basis-independent. The 
reduced matrix element in Eq.~(106) is identical with the one
introduced by Racah.$^7$ It is a rotational invariant that can 
be in general expressed in terms of basic invariants (e.g., reduced 
matrix element of Wigner  unit  operator, $\overline{W}$ and 
$X$ coefficients). Therefore, it does not depend 
on the labels $\alpha_1$, $\alpha_2$ and $\alpha$. On the contrary, 
the $\overline{f_r}$ coefficient in Eq.~(106), defined by Eq.~(79),  
depends on the labels $\alpha_1$, $\alpha_2$ and $\alpha$. The 
information on the geometry is entirely contained in the   
$\overline{f_r}$ coefficient.

\section{Concluding remarks}
The main results presented in this paper are the following. 
(i) The
nondeformed Lie algebra $\mbox{su}_{2}$ may be constructed 
from two commuting 
$q$-deformed oscillator algebras with $q$ being a root of unity; the latter  
oscillator algebras are associated with (truncated) harmonic oscillators having a
finite number of eigenvectors. 
(ii) This construction leads to the polar
decomposition of the generators $J_+$ and 
                                $J_-$ of SU(2) 
originally introduced by L\'evy-Leblond.$^{13}$ 
(iii) The familiar $\{ J^2, J_z \}$ quantization 
scheme with the (usual) standard spherical basis  
$\{ |jm \rangle : 2j \in {\bf N}, \ m = -j, -j+1, \cdots, j\}$, 
corresponding to
the canonical chain of groups SU(2)~$\supset$~U(1), is thus replaced by the 
$\{ J^2, U_r \}$ quantization scheme with a (new) basis, 
namely, the nonstandard basis    
$B_r 
= \{ |j \alpha ; r \rangle : 2j \in {\bf N}, 
\ \alpha = -jr, -jr+1, \cdots, -jr + 2j\}$. 
(iv) The Wigner-Racah algebra of  SU(2)  may be developed in the 
$\{ J^2, U_r \}$ scheme.  

These various results should be useful in problems 
involving axial symmetry and in the investigation 
of quantum mechanics on a finite Hilbert space 
as developed by several authors.$^{15}$ To make 
the latter point clear, let us write $S$ (see Eq.~(56)) 
and $B_r$ (see Eq.~(57)) as 
\begin{eqnarray}
S   = \bigcup_{j=0}^{\infty} s^j
\end{eqnarray}
and 
\begin{eqnarray}
B_r = \bigcup_{j=0}^{\infty} b_r^j
\end{eqnarray}
where $s^j$ and $b_r^j$ are two bases that span the 
subspace $\varepsilon(j)$. It is clear that $s^j$ 
and $b_r^j$ are two mutually unbiased bases (MUB's) 
in the sense that
\begin{eqnarray}
\vert \langle jm | j \alpha ; r \rangle \vert = 
{1 \over \sqrt{\dim {\varepsilon(j)}}}
\end{eqnarray}
It is known that the MUB's are especially useful 
in the theory of quantum information. In this respect, 
a connection between our results and some of the ones 
in Ref.~[15] is presently under study.

\section*{\bf Acknowledgments}
It is a real pleasure to dedicate this paper to Prof. Josef Paldus 
on the occasion of his 70th anniversary in recognition 
of his important contribution to various domains of the quantum 
theory of molecular electronic structure. The quantum chemistry 
community greatly benefits from his works on nonrelativistic and 
relativistic electronic systems via second quantization and field-theoretical 
approaches, diagrammatic methods, Lie-like and Clifford-like algebraic techniques. His 
numerous results on the many-electron correlation problem, on the configuration 
interaction method, on the 
coupled-cluster theory, and on density matrix calculations should 
be a source of inspiration for young theoretical chemists. The 
present author is very indebted to Prof. Paldus for interesting 
discussions and the kind hospitality extended to him on the occasion 
of several fruitful visits at the Mathematics Department of the 
University of Waterloo. {\em Merci beaucoup Jo}.


\newpage

\begin{thebibliography}{99}
\itemsep=-3pt

\bibitem{001} 
J. \v{C}\'{\i}\v{z}ek and J. Paldus, {\em Int. J. Quantum Chem.} 12, 875 (1977). 

J. \v{C}\'{\i}\v{z}ek, B.~G. Adams and J. Paldus, {\em Phys. Scr.} 21, 364 (1980).

\bibitem{002} 
M. Kibler and T. N\'egadi, {\em Lett. Nuovo Cimento} 37, 225 (1983); 
                           {\em J. Phys. A: Math. Gen.} 16, 4265 (1983); 
                           {\em Phys. Rev.} A 29, 2891 (1984). 

M.R. Kibler, {\em J. Mol. Phys.} 102, 1221 (2004).

\bibitem{003} 
J. Paldus, J. \v{C}\'{\i}\v{z}ek and I. Shavitt, {\em Phys. Rev. A} 5, 50 (1972).  

J. Paldus, {\em J. Chem. Phys.} 57,  638 (1974).

J. Paldus, {\em J. Chem. Phys.} 61, 5321 (1974).

J. Paldus, {\em Int. J. Quantum Chem. S} 9, 165 (1975).

J. Paldus and J. \v{C}\'{\i}\v{z}ek, {\em Adv. Quantum Chem.} 9, 105 (1975).

J. Paldus, {\em Phys. Rev. A} 14, 1620 (1976).

J. Paldus, {\em J. Chem. Phys.} 67, 303 (1977).

J. Paldus, B.~G. Adams and J. \v{C}\'{\i}\v{z}ek, 
{\em Int. J. Quantum Chem.} 11, 813 (1977). 

B.~G. Adams, J. Paldus and J. \v{C}\'{\i}\v{z}ek, 
{\em Int. J. Quantum Chem.} 11, 849 (1977). 

J. Paldus, J. \v{C}\'{\i}\v{z}ek, M. Saute and A. Laforgue, 
{\em Phys. Rev. A} 17, 805 (1978).

P.E.S. Wormer and J. Paldus, {\em Int. J. Quantum Chem.} 16, 1307 (1979).

J. Paldus and P.E.S. Wormer, {\em Int. J. Quantum Chem.} 16, 1321 (1979).

P.E.S. Wormer and J. Paldus, {\em Int. J. Quantum Chem. 18},  841 (1980).

J. Paldus and M.J. Boyle, {\em Int. J. Quantum Chem.} 22, 1281 (1982).

J. Paldus, M. Takahashi, and R.W.H. Cho, {\em Phys. Rev. B} 30, 4267 (1984).

J. Paldus and P. Piecuch, {\em Int. J. Quantum Chem.} 42, 135 (1992).

X. Li and J. Paldus, {\em J. Chem. Phys.} 119, 5334 (2003). 

J.L. Stuber and J. Paldus, in 
{\em Fundamental World of Quantum Chemistry}, Vol.~I, Eds. E.J. 
Br\"andas and E.S. Kryachko (Kluwer, Dordrecht, 2003). 

\bibitem{004} 
M.-l. Ge and G. Su, {\em J. Phys. A: Math. Gen.} 24, L721 (1991). 

M.A.~Mart\'\i n-Delgado, {\em J.~Phys.~A~: Math. Gen.} 24, L1285 (1991).  

C.R. Lee and J.-P. Yu, {\em Phys. Lett. A} 164, 164 (1992).

G. Su and M.-l. Ge, {\em Phys. Lett. A} 173, 17 (1993). 

J.A. Tuszy\'nski, J.L. Rubin, J. Meyer and M. Kibler, 
{\em Phys. Lett. A} 175, 173 (1993). 

V.I. Man'ko, G. Marmo, S. Solimeno and F. Zaccaria, 
{\em Phys. Lett. A} A176, 173 (1993).

R.-R. Hsu and C.-R. Lee, {\em Phys. Lett. A} 180, 314 (1993).

Ya.I. Granovskii and A.S. Zhedanov, {\em Mod. Phys. Lett. A} 8, 1029 (1993). 

M. Chaichian, R.G. Felipe and C. Montonen, {\em J. Phys. A: Math. Gen.} 26, 4017 (1993). 

R.K. Gupta, C.T. Bach and H. Rosu, {\em J. Phys. A: Math. Gen.} 27, 1427 (1994). 

M.A. R.-Monteiro, I. Roditi and L.M.C.S. Rodrigues, 
{\em Phys. Lett. A} 188, 11 (1994).

R.-S. Gong, {\em Phys. Lett. A} 199, 81 (1995). 
 
M. Daoud and M. Kibler, {\em Phys. Lett. A} 206, 13 (1995). 

\bibitem{005}  
E. Witten, {\em Nucl.~Phys.~B} 330, 285 (1990). 

S. Iwao, {\em Prog.~Theor.~Phys.} 83, 363 (1990).

D. Bonatsos, P.P. Raychev, R.P. Roussev and 
Yu.F. Smirnov, {\em Chem. Phys. Lett.} 175, 300 (1990). 
 
Z. Chang and H. Yan, {\em Phys.~Lett.~A} 154, 254 (1991). 

Z. Chang H.-Y. Guo and H. Yan, {\em Phys.~Lett.~A} 156, 192 (1991).

Z. Chang and H. Yan, {\em Phys.~Lett.~A} 158, 242 (1991). 

D. Bonatsos, S.B. Drenska, P.P. Raychev, R.P. Roussev and Yu.F. Smirnov, 
{\em J.~Phys.~G: Nucl. Part. Phys.} 17, L67 (1991). 

D. Bonatsos, P.P. Raychev and A. Faessler,  
{\em Chem.~Phys.~Lett.} 178, 221 (1991). 

D. Bonatsos, E.N. Argyres and P.P. Raychev, 
{\em J.~Phys.~A: Math. Gen.} 24, L403 (1991). 

L. Jenkovszky, M. Kibler and A. Mishchenko, {\em Mod. Phys. Lett A} 10, 51 (1995). 

R. Barbier and M. Kibler, in {\em Finite Dimensional Integrable Systems}, 
Eds. A.N. Sissakian and G.S. Pogosyan (JINR, Dubna, 1995). 

R. Barbier and M. Kibler, in {\em Modern Group Theoretical Methods in Physics}, Eds. J. 
Bertrand, M. Flato, J.-P. Gazeau, D. Sternheimer and M. Irac-Astaud
(Kluwer, Dordrecht, 1995). 

R. Barbier and M. Kibler, {\em Rep. Math. Phys.} 38, 221 (1996). 

\bibitem{006}  
E.P. Wigner, {\em Z. Phys.} 43, 624 (1927). 

E.P. Wigner, {\em Am. J. Math.} 63, 57 (1941). 

E.P. Wigner, {\em Group Theory and its Application to the Quantum Mechanics 
of Atomic Spectra} (Academic Press, New York, 1959). 

E.P. Wigner, in {\em Quantum Theory of Angular Momemtum}, Eds.~L.C. 
Biedenharn and H. van Dam (Academic Press, New York, 1965). 

E.P. Wigner, in {\em Spectroscopic and Group Theoretical Methods in Physics}, 
Eds.~F. Bloch, S.G. Cohen, A. de-Shalit, S. Sambursky and I. Talmi 
(North-Holland, Amsterdam, 1968). 

E.P. Wigner, {\em Proc. Roy. Soc. London} A 322, 181 (1971). 

F.G. Goldrich and E.P. Wigner, {\em Can. J. Math.} 24, 432 (1972).

E.P. Wigner, {\em SIAM J. Appl. Math.} 25, 169 (1973).

\bibitem{007}  
G. Racah, {\em Phys. Rev.} 61, 186  (1942).

G. Racah, {\em Phys. Rev.} 62, 438  (1942). 

G. Racah, {\em Phys. Rev.} 63, 367  (1943).

G. Racah, {\em Phys. Rev.} 76, 1352 (1949). 

G. Racah, {\em Phys. Rev.} 78, 622  (1950).

G. Racah, {\em Bull. Res. Council Isr., Sect.~F} 8, 1 (1959). 

G. Racah, in {\em Ergebnisse der exakten Naturwissenschaften} 
(Springer-Verlag, Berlin) 37, 27 (1965).

U. Fano and G. Racah, {\em Irreducible Tensorial Sets} 
(Academic Press, New York, 1959). 

G. Racah and J. Stein, {\em Phys. Rev.} 156, 58 (1967). 

\bibitem{008} 
G. Grenet and M. Kibler, {\em Phys. Lett. A} 68, 147 (1978).

G. Grenet and M. Kibler, {\em Phys. Lett. A} 71, 323 (1979).

M. Kibler and G. Grenet, {\em J. Math. Phys.} 21, 422 (1980).

\bibitem{009}
M. Kibler, {\em J. Mol. Spectrosc.}        26,  111 (1968); 
           {\em Int. J. Quantum Chem.}        3,  795 (1969); 
           {\em C.~R. Acad. Sci. (Paris) B} 268, 1221 (1969).

M.R. Kibler, {\em J. Math. Phys.}         17,  855 (1976); 
             {\em J. Mol. Spectrosc.}   62,  247 (1976); 
             {\em J. Phys. A: Math. Gen.} 10, 2041 (1977).

M.R. Kibler and P.A.M. Guichon, {\em Int. J. Quantum Chem.} 10, 87 (1976).

M.R. Kibler and G. Grenet, {\em Int. J. Quantum Chem.} 11, 359 (1977). 

M.R. Kibler, {\em Int. J. Quantum Chem.} 23, 115 (1983). 

\bibitem{010}
J. Moret-Bailly, {\em J. Mol. Spectrosc.} 15, 344 (1965). 

J.P. Champion, G. Pierre, F. Michelot and J. Moret-Bailly, 
{\em Can. J. Phys.} 55, 512 (1977). 

J. Patera and P. Winternitz, {\em J. Math. Phys.} 14, 1130 (1973). 

J. Patera and P. Winternitz, {\em J. Chem. Phys.} 65, 2725 (1976).  

L. Michel, in {\em Group Theoretical Methods in Physics}, 
Eds.~R.T. Sharp and B. Kolman (Academic Press, New York, 1977). 

\bibitem{011}
M.~Arik and D.D.~Coon, {\em J.~Math.~Phys.} 17, 524 (1976). 

L.C. Biedenharn, {\em J. Phys. A: Math. Gen.} 22, L873 (1989). 

C.-P. Sun and H.-C. Fu, {\em J. Phys. A: Math. Gen.} 22, L983 (1989).

A.J. Macfarlane, {\em J. Phys. A: Math. Gen.} 22, 4581 (1989). 

D.B. Fairlie, P. Fletcher and C.K. Zachos, {\em J. Math. Phys.} 31, 1088 (1990). 

O.W. Greenberg, {\em Phys. Rev. D} 43, 4111 (1991). 

G. Rideau, {\em Lett. Math. Phys.} 24, 147 (1992). 

M.R. Kibler, in {\em Symmetry and 
Structural Properties of Condensed Matter}, Eds.~W. Florek, D. Lipi\'nski 
and T. Lulek (World Scientific, Singapore, 1993). 

M. Daoud, Y. Hassouni and M. Kibler, in {\em Symmetries in Science X},
Eds.~B. Gruber and M. Ramek (Plenum Press, New York, 1998). 

M. Daoud and M. Kibler, {\em Phys. Lett. A} 321, 147 (2004). 

M. Kibler and M. Daoud, in {\em Fundamental World of Quantum Chemistry}, 
Vol.~III, 
Eds. E.J. Br\"andas and E.S. Kryachko (Kluwer, Dordrecht, 2003). 

\bibitem{012}
J. Schwinger, in {\em Quantum Theory of Angular Momemtum}, 
Eds.~L.C. Biedenharn and H. van Dam (Academic Press, New York, 1965). 

\bibitem{013} 
J.-M. L\'evy-Leblond, {\em Rev. Mex. F\'\i s.} 22, 15 (1973).

\bibitem{014} 
A.R. Edmonds, {\em Angular Momentum in Quantum Mechanics}
(Princeton University Press, Princeton, 1960).

\bibitem{015}
A. Vourdas, {\em Phys. Rev. A} 41, 1653 (1990). 

A. Vourdas, {\em Phys. Rev. A} 43, 1564 (1991).

A. Vourdas and C. Bendjaballah, 
            {\em Phys. Rev. A} 47, 3523 (1993).
	    
A. Vourdas, {\em J. Phys. A: Math. Gen.} 29, 4275 (1996).

M. Planat and H.C. Rosu, {\em Phys. Lett. A} 315, 1 (2003).  

M. Planat and H. Rosu, {\em J. Opt. B: Quantum Semiclassical Opt.} 6, S583 (2004). 
 
M. Saniga, M. Planat and H. Rosu, {\em J. Opt. B: Quantum Semiclassical Opt.} 6, L19 (2004). 

M. Saniga and M. Planat, arXiv: quant-ph/0409184. 

A.B. Klimov, L.L. S\'anchez-Soto and H. de Guise, arXiv: quant-ph/0410135.

A.B. Klimov, L.L. S\'anchez-Soto and H. de Guise, arXiv: quant-ph/0410155.

\end{thebibliography}
\end{document}